\documentclass[12pt]{article}
\begin{document}
\input amssym.def
\input amssym
\centerline{\bf Pure Representability Problem and New Models of }
\centerline{\bf the Electronic Fock Space }
\bigbreak
\centerline {A. I. Panin}
\bigbreak
\centerline{ Chemistry Department, St.-Petersburg State University,}
\centerline {University prospect 2, St.-Petersburg 198504, Russia }
\centerline { e-mail: andrej@AP2707.spb.edu }
\bigbreak

{\bf ABSTRACT: }{\small  New models of the Fock space sector corresponding to
some fixed number of electrons are introduced. These models originate from
the representability theory and their practical implementation
may lead to essential reduction of dimensions of intermediate CI spaces. A certain  zero-order theory
that gives  wavefunctions approximately equivalent to ones
obtained by accounting all excitations
from the Hartree-Fock reference state  up to the $q$-th order is
proposed. Simple numerical examples are given to illustrate our approach.
}


{\small electron correlation.}

\bigbreak
\centerline{\bf 1.Introduction}
\bigbreak

In present work new realizations of $p$-electron sector of the
Fock space are proposed and their simplest properties are studied.
Notations and terminilogy of this article are new for Quantum Chemistry
and require  therefore few preliminary comments. The basic idea of our
approach is  widespread in modern mathematics: to assemble information
about an  object as a whole using available local information. This idea
was properly formalized and systematically used by Jean Leray who introduced
his famous sheaves over
topological spaces and employed them in complex analysis
(see, e.g., \cite {Bredon}).
Since this branch of mathematics is not used in Quantum Chemistry , at least at
present, we found it possible to borrow existing mathematical terms ``sheaf''
and ``germ''
filling them with quite different meaning. Namely, we introduce
a family of $q$-electron functions (germs)
and call such family ``sheaf'' if the germs constituting this sheaf are
subject to  certain ``gluing'' conditions.  Then it is  shown how sheaves thus
defined can be used to restore unique $p$-electron wavefunction. The whole
theory was motivated by the theorem from the pure representability theory that was proved
in \cite{ Panin-1}. To make presentation of our approach more transparent and logical,
in the second section  necessary basic definitions and results from the
representability theory are formulated.
Our approach  is essentially
finite-dimensional and set-theoretical manipulations and technique of
enumerative combinatorics are extensively used.
For these reasons we
found it possible  to change notations that seems to be considered as
traditional in literature on the representability problem. Namely, capital
letters are normally used to denote subsets of the spin-orbital
index set $N$ and/or the orbital index set $M$. Number of elements (indices)
in index sets is usually denoted by lower case letters, e.g. $|N|=n$,
$|M|=m$, etc. With such notations it seems consistent to use lower case
letters for the number of electrons and for the current density matrix
order that are just the numbers of elements in the relevant spin-orbital
index sets. Throughout this article $n$ is the number of molecular spin orbitals (MSO), $m$ is
the number of molecular orbitals (MO), $p$ stands for the number of electrons, and $q$ is the
density matrix order.

In the third section  the formal theory of $(p,q)$-sheaves is presented.

In the fourth section  construction of correlated basis functions in configuration interaction
(CI) theory is discussed.

In the fifth section  it is demonstrated  how  the abstract mathematical theory of
$(p,q)$-sheaves   may be used for
replacing large-scale CI calculations by series of
CI calculations in relatively small dimensions.

\newpage

\centerline{\bf 2.Basic Definitions.}
\bigbreak

Let ${\cal F}_{n,1}$ be one-electron Fock space spanned by an orthonormal
set of $n$ molecular {\sl spin-orbitals\ }. Electronic Fock space
is defined as

$${\cal F}_n=\bigoplus\limits_{p=0}^n {\cal F}_{n,p} \eqno(1)$$

where

$${\cal F}_{n,p}=\bigwedge\limits^p{\cal F}_{n,1}\eqno(2)$$
$${\cal F}_{n,0}={\Bbb C} \eqno(3)$$
and  $\Bbb C$ is the field of complex numbers.

``Determinant'' basis vectors of the Fock space will be labelled
by subsets of the spin-orbital index set $N$: for any $R\subset N$ the
corresponding basis determinant will be denoted by $|R\rangle$. Note that
basis determinants are labelled by {\sl subsets}  and  all sign conventions
connected with their actual representation as  the Grassman
product of {\sl ordered} spin-orbitals are included in the definition of the
creation-annihilation operators:

$$a_i^{\dag}|R\rangle=(1-\delta_{i,R})(-1)^{\epsilon}|R\cup i\rangle
\eqno(4a)$$
$$a_i|R\rangle=\delta_{i,R}(-1)^{\epsilon}|R\backslash i\rangle\eqno(4b)$$

where

$$\delta_{i,R}=\cases{1,&if $i\in R$\cr
                      0,&if $i\notin R $\cr}\eqno(5)$$

may be considered as a possible generalization of the Kronecker $\delta$
symbol, and

$$\epsilon=|\{ 1,2,\ldots,i-1\} \cap R|\eqno(6)$$

is the sign counter.

Dirac's ket-bra realization of the  operator space  over
the electronic Fock
space  $(End_{\Bbb C}({\cal F}_n,{\cal F}_n)\sim {\cal F}_n\otimes {\cal F}_n^*)$
equipped   with the trace inner product
$$(z|t)= Tr(z^{\dag}t)\eqno(7)$$
will be used.

The contraction  operator over  ${\cal F}_n\otimes {\cal F}_n^*$ is defined as

$$c=\sum\limits_{i=1}^na_i\otimes a_i^{\dagger}\eqno(8)$$

Its restriction to the $p$-electron sector of the operator space is
identical (up to the nonessential combinatorial prefactor) to the commonly
used contraction operator \cite {Coleman-1,Coleman-2,Coleman-3,Coleman-4}.

Let $t_p\in {\cal F}_{n,p}\otimes {\cal F}_{n,p}^*$ be some p-electron
operator. It can be expanded via basis ``determinant generators'' as

$$t_p=\sum\limits_{R,S}^{(p)}t_{RS}|R\rangle \langle S|\eqno(9)$$

where the upper summation index $(p)$ indicates that the sum is taken over
all  $p$-element subsets of the spin-orbital index set $N$.  It can be
shown \cite {Panin-1,Panin-2} that

$$c^k(t_p)=k!\sum\limits_{R,S}^{(p)}t_{RS}\sum\limits_{K\subset R\cap S}
^{(k)}(-1)^{|(R\Delta S)\cap \Delta_K|}|R\backslash K\rangle\langle S
\backslash K|\eqno(10)$$

The set-theoretical operation ${\Delta}_K$ was introduced in \cite{ Panin-2} and its
main properties were investigated in \cite {Panin-1} but due to its importance for
the {\sl uniform} treatment of the sign prefactors arising in the expressions of
the type of Eq.(10) we repeat here necessary definitions.

For two arbitrary
index sets $R$ and $S$ their {\sl symmetric difference} is $R\Delta S=
(R\cup S)\backslash (R\cap S)$. This binary operation on the set of all subsets
of some given set is commutative and associative with the empty set as its unit.
For any subset $K=\{k_1,k_2,\ldots,k_l\}\subset N$ we put
$${\Delta}_K=\{1,2,\ldots ,k_1\}\Delta \{1,2,\ldots ,k_2\}\Delta
\ldots \Delta \{1,2,\ldots ,k_l\}.\eqno(11)$$
For example, if $K=\{2,4,5\}$ in $N=\{1,2,3,4,5,\ldots\}$ then
$${\Delta}_K=\{1,2\}\Delta \{1,2,3,4\}\Delta \{1,2,3,4,5\}
=\{3,4\}\Delta \{1,2,3,4,5\}=\{1,2,5\}$$

The electronic Hamiltonian associated with some chosen  spin-orbital basis set in
one-electron Fock space is of the form

$$H=\sum\limits_{i,j}\langle i|h|j\rangle a_i^{\dag}a_j+
{1\over 2}\sum\limits_{i,j,k,l}\langle ij|{1\over{r_{12}}}|kl\rangle
a_i^{\dag}a_j^{\dag}a_la_k\eqno(12)$$

The energy functional on the Fock space  is defined as

$$E(t)=Tr(Ht)\eqno(13)$$

and is a linear mapping from ${\cal F}_n\otimes {\cal F}_n^*$ to
the field of complex numbers.

Using specific form of the electronic
Hamiltonian, it is possible to contract the energy domain and redefine the
energy in terms of $q$-electron operators $(q\ge 2)$.  Introducing reduced
Hamiltonian that acts on the $q$-electron sector of the Fock space

$$H_{p\to q}=\frac{q-1}{p-1}\sum\limits_{i,j}\langle i|h|j
\rangle a_i^{\dag}a_j+
{1\over 2}\sum\limits_{i,j,k,l}\langle ij|{1\over{r_{12}}}|kl\rangle
a_i^{\dag}a_j^{\dag}a_la_k\eqno(14)$$

we can recast the energy expression (13) in the form

$$E(t_p)=\frac{{p\choose 2}}{{q\choose 2}}
Tr\left [H_{p\to q}\frac{q!}{p!}c^{p-q}(t_p)\right ]
\eqno(15)$$
where $t_p$ is some $p$-electron operator. On the right hand side of  this
equation $q$-electron image $(2\le q\le min(p,n-p))$ of $t_p$ with respect to
the contraction appears.

Of actual interest in Quantum Chemistry are the so-called pure $p$-electron
states that are associated with $p$-electron wavefunctions. The corresponding
pure representability problem was formulated in the very first articles
of Coleman \cite {Coleman-1,Coleman-2,Coleman-3,Coleman-4}:

(PR) {\sl To find the necessary and sufficient conditions that should be imposed
on $q$-electron operator $t_q$ to guarantee that there exists at least one
$p$-electron wave function $\Psi$ such that
$ \frac{q!}{p!}c^{p-q}|\Psi \rangle \langle \Psi|=t_q$.}

From the very beginning it was understood that the structure of the set of all
pure representable $q$-electron operators is very complicated and that
its constructive description that could be of any practical value hardly
exists at all. The well-known theorem of the Convex Sets Theory states that a real-valued
linear function defined on some compact convex set necessarily reaches its minimal
value at one of the extreme points of this function domain. Therefore, if
the set of all pure representable $q$-electron operators is replaced by its convex
hull, the set of certainly more simple structure will be obtained and  one can guarantee
that the direct search of energy minimal value  on this set will lead  to the optimal pure state.
This was also employed by Coleman who formulated his famous ensemble representability
problem:

(ER) {\sl To find the necessary and sufficient conditions that should be imposed
on $q$-electron operator $t_q$ to guarantee that there exists
ensemble of $p$-electron states ${\Psi}_i$ such that
$ \frac{q!}{p!}c^{p-q}\sum\limits_i{\lambda}_i|{\Psi}_i
\rangle \langle {\Psi}_i|=t_q$. }

Here $\sum\limits_i{\lambda}_i=1$ and ${\lambda }_i\ge 0$.

It is very easy to find three very simple conditions that should
necessarily be satisfied by any representable $q$-electron operator:

(i) Nonnegativeness;

(ii) Hermiteancy;

(iii) Fixed trace value (usually $Tr(t_q)=1$).

The set of all $q$-electron  operators satisfying these three conditions
will be denoted by ${\cal E}_{n,q}$. Operators from  ${\cal E}_{n,q}$ are called
{\sl density operators}.

In his pioneer works Coleman \cite {Coleman-1,Coleman-2,Coleman-3,Coleman-4} found the solution of the ensemble representability
problem for the case $q=1$ and formulated two very strong necessary conditions
for the case $q=2$.

In our previous work \cite {Panin-1} we generalized part of Coleman's results to treat the case of
density operators of arbitrary order  and found the following description for
the set ${\cal V}_{n,p,q}$ giving an exterior approximation of the set of all
ensemble representable density operators of order $q$:
$${\cal V}_{n,p,q}=A^{-1}(n,p,q){\cal E}_{n,q}\cap {\cal
E}_{n,q}\eqno(16)$$
where the automorphism $A(n,p,q)$ is determined by its matrix
representation as
$$A(n,p,q)e^{IJ}_K=
(-1)^{\alpha_{IJ}} \frac{{p\choose q}}{{p-s\choose q-s}{n-p\choose q}}
\sum\limits_{K'\subset N\backslash (I\cup J)}^{(q-s)}(-1)^{|K\cap K'|}
\frac{{p+|K\cap K'|-q-1\choose {|K\cap K'|}}}{{q-s\choose |K\cap K'|}}
e^{IJ}_{K'}\eqno(17)$$
with respect to specially selected operator basis
$$e_L^{IJ}=(-1)^{|(I\cup J)\cap \Delta_L|}|I\cup L\rangle \langle J\cup L|
\eqno(18)$$
Here $I\cap J=\emptyset $, $s=|I|=|J|$,  and
$\alpha_{IJ}=|(I\cup J)\cap \Delta_{N\backslash (I\cup J)}|$.
In the same work very simple connection between
the operator $A(n,p,q)$ and its inverse was found:
$$A^{-1}(n,p,q)=A(n,n-p,q)\eqno(19)$$
Let us  apply the operator $A(n,p,q)\frac{q!}{p!}c^{p-q}$ to some pure
state $|\Psi \rangle \langle \Psi|$ where
$$\Psi=\sum\limits_{R\subset N}^{(p)}C_R|R\rangle\eqno(20)$$
Rather complicated manipulations (see Appendix D of \cite {Panin-1}) lead to
$$A(n,p,q)\frac{q!}{p!}c^{p-q}|\Psi\rangle\langle \Psi|=
\frac{1}{{n-p\choose q}}\sum\limits_{Z\subset N}^{(p+q)}d_q(Z)\eqno(21)$$
where
$$d_q(Z)=|\psi_Z\rangle \langle\psi_Z|\eqno(22)$$
and
$$|\psi_Z\rangle =\sum\limits_{S\subset Z}^{(q)}(-1)^{|(Z\backslash S)
\cap \Delta_Z|}C_{Z\backslash S}|S\rangle \eqno(23)$$
Let us introduce the set
$$B_{n,p,q}=\{(Z,S)\subset N\times N:|Z|=p+q \& |S|=q \& S\subset Z\}\eqno(24)$$
and the equivalence relation on this set
$$ (Z,S) \sim (Z',S') \Leftrightarrow Z\backslash S=Z'\backslash S' \eqno(25)$$
The set of all equivalence classes ${\overline B}_{n,p,q}$ contains exactly $n\choose p$
elements and in each equivalence class $\overline {(Z,S)}$ there are $n-p\choose q$
elements.

Eqs.(21)-(23) together with the definition (24) make it possible to formulate
the following necessary and sufficient conditions of pure representability.
{\bf Theorem . } {\sl $q$-electron operator $t_q$ is representable by pure
$p$-electron state if and only if its image $d_q$ with respect to
$A(n,p,q)$ satisfies the following conditions:

(i) There exists expansion
$$d_q=\frac{1}{{n-p\choose q}}\sum\limits_{Z\subset N}^{(p+q)}d_q(Z)\eqno(26)$$
such that for each $Z\subset N$, $d_q(Z)$ is either the null operator or
corresponds to (unnormalized) pure $q$-electron state $|\psi_Z\rangle$;

(ii) The mapping
$$ (Z,S) \to (-1)^{|(Z\backslash S)\cap \Delta_Z|}
\langle S|\psi_Z\rangle \eqno(27)$$
is constant on the equivalence classes $\overline {(Z,S)}$.}

Unfortunately, condition (i) of the above Theorem is depressingly
non-constructive. Only for the case  $p+q=n$ conditions (i) and (ii)
are trivial and the following corollary of the main theorem may be
formulated.

{\bf Corollary 1}. {\sl For $p+q=n$,  $q$-electron operator $t_q$ is representable by pure
$p$-electron state if and only if its image $d_q$ with respect to
$A(n,p,q)$ corresponds to pure $q$-electron state.}

The next corollary shows how to construct $p$-electron wavefunction
from q-electron ones.

{\bf Corollary 2} {\sl If the conditions of  Theorem are fullfilled then (up to
normalization) the required pure state may be presented as
$$ |\Psi\rangle=\sum\limits_{\overline {(Z,S)}}(-1)^
{|(Z\backslash S)\cap \Delta_Z|}\langle S|\psi_Z\rangle
|Z\backslash S\rangle \eqno(28)$$
where the sum runs over equivalence classes of the set $B_{n,p,q}$ modulo the
equivalence relation (25).}

The above Theorem was first formulated and proved in our previous article \cite {Panin-1}.
Its Corollary 1 has analogue in the ensemble representability theory where the
constructive description of the set of all representable density operators
of order $q$ for the case $n=p+q$ was well-known for years \cite {Erdahl}.

At first glance the above Theorem  is hardly of any practical value since in general
case we do not have even the slightest idea how to expand density operator $d_q$.
In the next section it will be  shown that it is a wrong impression and that this Theorem
opens a completely new direction  both in mathematical and computational
chemistry.

\newpage
\centerline{\bf 3. Representation of Many Electron Wave Function As a   }
\centerline {\bf Sheaf of q-Electron Germs.}
\bigbreak
For any $(p+q)$-element subset $Z$ of the spin-orbital index set $N$ we put
$${\cal F}_{n,q}(Z)=\bigoplus\limits_{S\subset Z}^{(q)}{\Bbb C}|S\rangle
\eqno(29)$$
{\bf Definition.} Family $\{{\psi}_Z\}_{Z\subset N}$ of $q$-electron
functions, ${\psi}_Z\in {\cal F}_{n,q}(Z)$, is called a sheaf of q-electron
germs of $p$-electron wavefunction, or just a $(p,q)$-sheaf, if the mapping
$$ (Z,S) \to (-1)^{|(Z\backslash S)\cap \Delta_Z|}\langle S|{\psi}_Z\rangle,
|Z|=p+q,|S|=q,S\subset Z\eqno(30)$$
is constant on the equivalence classes of the set $B_{n,p,q}$ modulo the
equivalence relation (25).

The set of all $(p,q)$-sheaves will be denoted as ${\cal S}_{n,p,q}$.

The Theorem formulated in the preceding section and its Corollary 2 lead
to the conclusion that there exists one-to-one correspondence between
$p$-electron wavefunctions
\footnote {By an abuse of terminology  arbitrary vector of $p$-electron
sector of the Fock space is called wave function even if it is not normalized.}
and $(p,q)$-sheaves for each fixed $q$. Indeed,
arbitrary wavefunction $\Psi$ represented by its Full CI (FCI) expansion (20)
is corresponded with  $(p,q)$-sheaf
$$s_{n,p,q} : \Psi \to \{{\psi}_Z\}_{Z\subset N}\eqno(31)$$
where $q$-electron wavefunctions ${\psi}_Z$ are given by Eq.(23). On the
other hand, due to the  conditions of gluing (30) each $(p,q)$-sheaf
uniquely determines $p$-electron wavefunction in accordance to Eq.(28).
The mapping $s_{n,p,q}$ performs disassembling of $p$-electron wavefunction into the
family of $q$-electron ones whereas its inverse assembles $p$-electron function
from its $q$-electron germs.

Disassembling mapping (31) may be used to transfer all relevant structures from
the Fock space ${\cal F}_{n,p}$ to the set ${\cal S}_{n,p,q}$. The most
important {\sl linear structure} is transfered as
$$s_{n,p,q}(\sum\limits_i c_i{\Psi}^i)=\left \{
(\sum\limits_i c_i{\Psi}^i)_Z \right\}_{Z\subset N}=
\sum\limits_i c_i\{ {\psi}^i_Z\}_{Z\subset N}\eqno(32)$$
whereas {\sl the inner product (Euclidean structure)} is defined as
$$\langle \{{\psi}_Z\}_{Z\subset N}|\{{\phi}_Z\}_{Z\subset N}\rangle
=\frac{1}{{{n-p}\choose q}}
\sum\limits_{Z\subset N}^{(p+q)}\langle {\psi}_Z|{\phi}_Z\rangle \eqno(33)$$
To avoid cumbersome notations, we agree to omit, when possible,
subscript $Z\subset N$ in writing elements from ${\cal S}_{n,p,q}$.

With the aid of the disassembling mapping (31) it is possible {\sl for each fixed} $q$
to identify the Fock space ${\cal F}_{n,p}$ with the space of all $(p,q)$-sheaves
${\cal S}_{n,p,q}$ considering the later just as an isomorphic model of the FCI
space.

The mapping
$${\pi}_{n,p,q}(Z):\{ {\psi}_{Z'}\}_{Z'\subset N} \to {\psi}_Z\eqno(34)$$
is obviously linear and performs a {\sl projection} of the vector space  ${\cal S}_{n,p,q}$
on the vector space ${\cal F}_{n,q}(Z)$. Among its sections (right inverses) there
exists in a certain sense canonical one that is defined as follows. To each
$q$-electron wavefunction
$${\psi}_Z=\sum\limits_{S\subset  Z}^{(q)}C_S|S\rangle =
\sum\limits_{R\subset  Z}^{(p)}(-1)^{|R\cap {\Delta}_Z|}{\bar C}_R|Z\backslash R\rangle
\eqno(35)$$
where  $R=Z\backslash S$ and
$${\bar C}_R=(-1)^{|R\cap {\Delta}_Z|}C_S\eqno(36)$$
it is possible to put into correspondence the unique $p$-electron wavefunction
$${\Psi}_Z=\sum\limits_{R\subset Z}^{(p)}{\bar C}_R|R\rangle \eqno(37)$$
with its subsequent disassembling. The resulting section of the projection
(34) will be denoted by $j_{n,p,q}(Z)$ and is a {\sl linear injective mapping}
satisfying the standard relation
$$ {\pi}_{n,p,q}(Z)\circ j_{n,p,q}(Z)=id_{{\cal F}_{n,q}(Z)}\eqno(38)$$
where $id_{{\cal F}_{n,q}(Z)}$ is the identity operator over ${\cal F}_{n,q}(Z)$.
As follows from this equation, the image of a certain $q$-electron wavefunction
${\psi}_Z$ with respect to $j_{n,p,q}(Z)$, that will be denoted as $\{{\psi}_{ZZ'}\}_{Z'\subset N}$ ,
posesses the  property  :
$${\psi}_Z={\psi}_{ZZ}\eqno(39)$$
or, in the other words, the initial $q$-electron wavefunction ${\psi}_Z$
is necessarily a germ of the $(p,q)$-sheaf $\{{\psi}_{ZZ'}\}_{Z'\subset N}$.
This property motivates the following definition.

{\bf Definition.} {\sl $(p,q)$-sheaf $\{{\psi}_Z\}_{Z\subset N}$ is called simple
if it is generated by one of its germ, that is, if there exists $(p+q)$-
element subset $Z'\subset N$ such that  }
$$\{{\psi}_Z\}_{Z\subset N} =j_{n,p,q}(Z')({\psi}_{Z'})\eqno(40)$$
{\bf Proposition 1.} {\sl $(p,q)$-sheaf generated by any its nonzero germ
corresponds to single determinant wavefunction.}

{\bf Proof.} If sheaf  $\{{\psi}_Z\}_{Z\subset N}$ is generated by any its
germ then the  wavefunction (37) does not depend on concrete subset $Z$.
If we suppose that it involves different determinants $|R\rangle$ and
$|R'\rangle $ in its expansion, then it is possible to find $(p+q)$-element subsets
$Z,Z'\subset N$ that separate $R$ and $R'$, that is $R\subset Z$,
$R'\subset Z'$,but $R\not\subset Z'$ and $R'\not\subset Z$.
In this case, however, ${\Psi}_Z$ can not be equal to ${\Psi}_{Z'}\blacksquare$

{\bf Definition.} {\sl $(p,q)$-sheaves generated by any its nonzero germ
are called  determinant sheaves.}

It is clear that determinant sheaves constitute an orthonormal basis of the
vector space ${\cal S}_{n,p,q}$. Selection of this basis will lead to
the approach completely equivalent to the FCI one in the basis of
$p$-electron determinants. All possible advantages and disadvantages of our
model for FCI space may appear in this case either  on the matrix element
evaluation level or on the level of  algorithms for solving partial
eigenvalue problem. Concrete formulas for matrix elements ready for practical
implementation  of the conventional CI method  are
collected in Appendix A. Here more general approach that is not
based on the determinant sheaves is discussed.

$q$-electron density operator $d_q$ of the form of Eq.(26)
is associated with each $(p,q)$-sheaf  $\{{\psi}_Z\}_{Z\subset N}$.
Recalling that $d_q$ originates from some $q$-electron density operator
$t_q=A^{-1}(n,p,q)d_q$ and that the operator $A(n,p,q)$ is Hermitean in trace
inner product, we can recast general energy expression (15) in the form
$$E(\{ {\psi}_Z\})=
Tr\left[\left (A^{-1}(n,p,q)H_{p\to q}\right )d_q\right] =$$
$$\frac{1}{{{n-p}\choose q}}\sum\limits_{Z\subset N}^{(p+q)}
\langle {\psi}_Z|A^{-1}(n,p,q)H_{p\to q}|
{\psi}_Z\rangle\eqno(41)$$
Note that in this formula the
initial definition of the reduced  Hamiltonian is slightly modified and the
current operator $H_{p\to q}$ differs from  that in Eq.(14) by
the combinatorial prefactor $\frac{{p\choose 2}}{{q\choose 2}}$.

In exactly the same manner general expression for the matrix elements of
the Hamiltonian between two $(p,q)$-sheaves may be written
$$H(\{ {\psi}_Z\},\{ {\phi}_Z\})=
\frac{1}{{{n-p}\choose q}}\sum\limits_{Z\subset N}^{(p+q)}
\langle {\psi}_Z|A^{-1}(n,p,q)H_{p\to q}|
{\phi}_{Z}\rangle\eqno(42)$$
{\sl This completely general formula expresses arbitrary matrix element
of $p$-electron Hamiltonian via matrix elements of the transformed
$q$-electron reduced Hamiltonian.} It is necessary to emphasize that
Eq.(42) does not couple $q$-electron spaces ${\cal F}_{n,q}(Z)$ with
different $Z$.

Of course, for their  practical use  it is necessary to
rewrite Eqs.(41)-(42) in terms of molecular orbitals taking into account
simplest spin symmetry restrictions. Relevant formulas are collected
in Appendix A.

Let us consider some arbitrary $(p,q)$-sheaf $\{{\psi}_Z\}_{Z\subset N}$.
Each its nonzero germ
${\psi}_{Z_j}$ generates  simple $(p,q)$-sheaf $\{{\psi}_{Z_jZ}\}_{Z\subset N}=
j_{n,p,q}(Z_j)({\psi}_{Z_j})$. The set of all such simple  sheaves will be
denoted as $J_{n,p,q}(\{{\psi}_Z\}_{Z\subset N})$ or
$J_{n,p,q}(\Psi)$ where $\Psi$ is the wave function corresponding to the initial
$(p,q)$-sheaf. The subspace of the vector space ${\cal S}_{n,p,q}$ generated by
the set $J_{n,p,q}(\Psi)$ will be called the $q$-subspace associated with
the wave function $\Psi$ (or the corresponding $(p,q)$-sheaf) and will be
denoted  as  $W_{n,p,q}(\Psi)$:
$$ W_{n,p,q}(\Psi )=\sum\limits_j{\Bbb C}\{{\psi}_{Z_jZ}\}_{Z\subset N}
\eqno(43)$$
It is  pertinent to note that the sum on the right-hand side of Eq.(43) is not
direct.

The importance of the definition (43) is due to the following fact.

{\bf Proposition 2.} {\sl $\{{\psi}_Z\}_{Z\subset N} \in  W_{n,p,q}(\Psi )$
for any $(p,q)$-sheaf $\{{\psi}_Z\}_{Z\subset N}=
s_{n,p,q}(\Psi).$ }

{\bf Proof.} Eq.(28) may be recast as
$$ |\Psi\rangle=\frac{1}{{{n-p}\choose q}}\sum\limits_{Z \subset N}
^{(p+q)}\sum\limits_{R\subset Z}^{(p)}(-1)^
{|R\cap \Delta_Z|}\langle Z\backslash R|\psi_Z\rangle
|R \rangle $$
From Eqs.(35)-(37) it follows that
$$ s_{n,p,q}\left (\sum\limits_{R\subset Z}^{(p)}(-1)^
{|R\cap \Delta_Z|}\langle Z\backslash R|\psi_Z\rangle|R\rangle \right )=
j_{n,p,q}(Z)({\psi}_Z)$$
But $j_{n,p,q}(Z)({\psi}_Z)$ is, by definition, a simple $(p,q)$-sheaf generated
by ${\psi}_Z$. As a result, $s_{n,p,q}(\Psi )$ is a linear combination of
its simple sheaves and, consequently, belongs to
$W_{n,p,q}(\Psi )$ $\blacksquare$

Thus, to each $p$-electron wavefunction it is possible to put into correspondence the
family of subspaces
$$\Psi \to \left ( W_{n,p,1}(\Psi), W_{n,p,2}(\Psi),
\ldots, W_{n,p,q_{max}}(\Psi)\right )\eqno(44)$$
and the associated  integral vector
$$ \Psi \to ind(\Psi )=(ind_1(\Psi ),ind_2(\Psi ),\ldots ,ind_{q_{max}}(\Psi ))\eqno(45)$$
where $s_{n,p,q}$ is the disassembling mapping (31), $q_{max}=min(p,n-p)$, and
$$ind_q(\Psi )=dim_{\Bbb C}W_{n,p,q}(\Psi)=rank_{\Bbb C}J_{n,p,q}(\Psi)\eqno(46)$$

{\bf Definition.}{\sl \quad  $ind_q(\Psi )$ is called the
$q$-index of the wavefunction $\Psi$ (or the corresponding $(p,q)$-sheaf).}

At present we do not know if indices (46) can be considered as integral characteristics of
the associated wave function. It is certainly true for the so-called {\sl formal Fock
space}.

{\bf Definition.} {\sl Vector space of all formal linear combinations of all
subsets of the index set N with complex coefficients is called formal Fock
space.}

Formal Fock space is in a certain sense universal model of the electronic
Fock space  with  orbital specific  erased. In this space
vector (45)  is the vector of integral chracteristics of formal $p$-electron
wavefunction $\Psi$.

On the Fock space additional structure connected with the action of the group
of unitary transformations $\bigwedge\limits ^p u$ induced by one-electron ones
appears. The same wavefunction has, in general, different CI coefficients in
different MSO bases and, consequently, is corresponded with different sheaves.
If the $q$- subspaces associated with these sheaves can be of different
dimensions  is the question to be  investigated.

When $q$ is fixed, then the index (46) may be  called just {\sl  the   CI
index}  of the wave function under consideration. The meaning of this index
is the following: for arbitrary wave function $\Psi$ the length of its
expansion in the subspace $W_{n,p,q}(\Psi)$ is not greater than
$ind_q(\Psi )$ (for fixed MSO basis set).

Let us consider the group $G_{n,p,q}(\Psi)$ of automorphisms of
the vector space $W_{n,p,q}(\Psi)$
leaving invariant the finite set of simple sheaves $J_{n,p,q}(\Psi)$.
Since $J_{n,p,q}(\Psi)$ generates $W_{n,p,q}(\Psi)$ this
group is isomorphic to the subgroup  of the symmetric group permuting
simple sheaves from $J_{n,p,q}(\Psi)$, and, consequently, is finite.
If for a given wave function $\Psi$ there exist non-trivial groups
$G_{n,p,q}(\Psi)$ is the problem for future study. At present
we can state only that
for each $q$ one-dimensional subspace of $W_{n,p,q}(\Psi)$ spanned by
$s_{n,p,q}(\Psi)$ carries trivial representation of $G_{n,p,q}(\Psi)$
(see Eq.(28)).

{\bf Definition.} {\sl $(p,q)$-sheaf $\{{\psi}_Z\}=s_{n,p,q}(\Psi)$ is called
stable
if it coincides with the lowest eigensheaf of the CI matrix in the subspace
$W_{n,p,q}(\Psi)$.}

It is easy to see that any sheaf can be transformed to the stable one in no more
than one step.

{\bf Definition.} {\sl Wavefunction $\Psi$ is called stable if
its $(p,q)$-sheaves $s_{n,p,q}(\Psi)$ are stable for each $2\le q\le min(p,n-p)$.}

All determinant functions   and  FCI lowest eigenvector are
necessarily stable. In general case with each $p$-electron wavefunction from
${\cal F}_{n,p}$ we can
associate the unique family of its stable images ${\Psi}_q$ from
${\cal S}_{n,p,q}$ $(q\ge 2)$ and the following inequalities always keep true:$\langle \Psi |H|\Psi \rangle \ge
\langle {\Psi}_q |H|{\Psi}_q \rangle$.

\newpage
\centerline{\bf 4. Construction of Correlated $q$-Electron Basis Functions.}
\bigbreak

It is well-known that the  proper selection of many electron basis
functions becomes the crucial  stage of the CI approach when the FCI
limit is unreachable. The commonly used methods  of basis functions
construction  employ very robust physical idea of
successive generation of k-fold excitations from some reference state
($k=1,2,\ldots$). In
the simplest case  the Hartree-Fock (HF) single determinant is used as a reference state.
We start our analysis with
the conventional CI approach in the basis of the determinant
$(p,q)$-sheaves.

Selecting in the subspace ${\cal F}_{n,q}(Z)$ the function
$${\psi}_Z={(-1)}^{|(Z\backslash S)\cap {\Delta}_Z|}|S\rangle $$
and applying the sheaf generation scheme described by Eqs.(35)-(37) lead to
the determinant $(p,q)$-sheaf
$$(Z',S') \to \left \{(-1)^{|R\cap {\Delta }_Z'|}|Z'\backslash R\rangle \right \}
_{Z'\supset R}\eqno(47)$$
where $R=Z\backslash S$, and it is immediately clear that for any pair $(Z',S')\in \overline {(Z,S)}$
the same determinant sheaf will be obtained.  This result can be reformulated in
the following manner: determinant $(p,q)$-sheaves are in one-to-one
correspondence with the  equivalence classes of the set $B_{n,p,q}$
modulo the equivalence relation (25). Selecting  HF determinant as the reference state,
it is easy to construct basis determinant sheaves by applying  Eq.(47) succesively
to the``excited pairs'' $(Z,S) \leftrightarrow Z\backslash S=R\backslash I\cup J$.
This at first glance  trivial scheme can be generalized in a very
interesting manner.

Let us try to select $q$-electron wavefunctions ${\psi}_Z\in {\cal F}_{n,q}(Z)$
with the following property: the corresponding simple sheaf should account as much
of correlation effects as possible. To this end we should
turn to Corollary 1 of our Theorem from  the second section. From this
corollary it follows that for fixed $Z$ the electronic energy minimum
is reached on the sheaf generated by the lowest eigenvector of the operator
$$A^{-1}(p+q,p,q)({\cal P}_ZH_{p\to q}{\cal P}_Z)\eqno(48)$$
where ${\cal P}_Z:{\cal F}_{n,q} \to {\cal F}_{n,q}(Z)$ is the standard  projection.

Moving along some equivalence class $\overline {(Z,S)}$, one can select  the lowest
eigenvector ${\psi}_{(Z,S)}$ of the operator (48)  obeying if necessary  some  additional
conditions of the type $\langle {\psi}_{(Z,S)}|S\rangle \ne 0$. Then all simple
$(p,q)$-sheaves $\left \{ {\psi}_{(Z,S)Z'}\right\}_{Z'\subset N}$ are to be constructed
and the
lowest eigenvector of the CI matrix in the vector space generated by these
sheaves should be found.
The resulting sheaf should be made stable by applying the routine described in
preceding section. With such an approach to each equivalence class
$\overline {(Z,S)}$ some vector accounting part of
the correlation effects is put into correspondence.
These (in general, non-orthogonal) vectors are in one-to-one correspondence
with $p$-electron determinants but the problem of their linear independence
is still to be studied.

Applying this scheme to the equivalence class of the HF
$p$-electron determinant leads to a certain zero-order theory
corresponding approximately to the conventional CI method accounting
excitations up to the q-th order. The main difference between the standard
CI approach  and  our scheme is in replacing large CI matrix partial diagonalization
by  a series of partial diagonalizations of relatively small orders. Indeed, for
class representatives matrices  of  one of the following orders arise:
${{p_{\alpha}+k} \choose  k}{{p_{\beta}+q-k}\choose {q-k}}$, where $k=0,1,\ldots ,q$.
After finding
optimal vector for each class representative,  it is necessary to solve partial diagonalization problem of
no more that  ${2m-p}\choose q$ order to  get equivalence class energy and wavefunction.
And the last diagonalization to stabilize the equivalence class wavefunction is
performed in the space of the minimal possible for a given wavefunction
dimension equal to its  CI index.   Numerical tests are discussed
in detail in the next section.  Here  simple comparison of relevant
dimensions is performed. For the fixed value $M_S$ of the total spin projection the number of
$\alpha$ electrons $p_{\alpha}$ and $\beta$ electrons $p_{\beta}$ is fixed and
the total number of all excitations up to the q-th order from the HF reference state
in  space of $m$ molecular orbitals is equal to
$$l_{(1,2,\ldots ,q)}=\sum\limits_{k=1}^q\sum\limits_{i=0}{{p_{\alpha}}\choose i}
{{m-p_{\alpha}}\choose i}{{p_{\beta}}\choose {k-i}}{{m-p_{\beta}}\choose {k-i}}
\eqno(50)$$
For the case of 80 electrons correlated in the one-electron space of 200 MOs,
$q=2$, and the total spin projection $M_S=0$ there are
$l_{(1,2)}=60 816 000$ excited configurations.
At the same time for each representative of the HF determinant
equivalence class   partial
eigenvalue problem of order either 1681 or 861 should be handled.
To get the wavefunction of the
HF equivalence class it is necessary to find the lowest eigenvalue of the matrix of
order not larger than 51040.

To improve the approximation described  one can try to
account relevant excitations from the zero-order wavefunction.
We believe, however, that it is possible to
develop very efficient new algorithms based on our model of the CI space.
Indeed, for some current $(p,q)$-sheaf $\{{\psi}_Z\}_{Z\subset N}$
approximate Newton-Raphson (NR) shift vector can be calculated with its
subsequent disassembling it to give NR sheaf $\{{\phi}_Z\}_{Z\subset N}$.
Then in the vector space generated by simple sheaves $\{{\psi}_{ZZ'}\}_{Z'\subset N}$
and $\{{\phi}_{ZZ'}\}_{Z'\subset N}$ the  CI problem should be solved to
to give new stable sheaf, etc. With such an approach
both $p$- and $q$-electron functions are actually optimized
and it can be expected that the final FCI solution will be obtained
by subsequent passing through  CI subspaces of moderate sizes.

\newpage
\centerline{\bf 5. Simple Numerical Examples.}
\bigbreak
To illustrate our approach, we performed very simple calculations of the HF
equivalence class wavefunctions for  two small molecules in  STO-6G
Gaussian basis set.
There are two reasons why  such primitive
examples were chosen. Firstly, the whole approach is just at its very beginning and
separate probably very complicated and time consuming  work should be done
to develop efficient computer schemes and create
the corresponding computer code. Secondly, for small molecules it is easy  to write
down explicitely all necessary equivalence classes, their energies, etc., to
make understanding of previous sections of this article easier to the reader
who has only vague impression about equivalence classes and is not
experienced in set-theoretical manipulations. In all calculations
Handy's split determinant representation was used \cite {Handy}. In this
representation  spin-orbital index set is replaced
by a pair of orbital index sets for $\alpha$ and $\beta $ spins.
Binary codes for index sets identification were also employed: e.g., index set $\{1,3,5\}$ as a subset
of the orbital index set $M=\{1,2,3,4,5,6,7\}$ is represented as a binary code
1010100. All RHF calculations to generate MO basis set were performed with
the aid of the GAMESS program \cite {GAMESS}.

From the definition (24) of the set $B_{n,p,q}$ and the equivalence
relation (25) it is easy to see that the same $(p+q)$-element subset
$Z\subset N$ occurs in exactly ${p+q}\choose q$ different equivalence classes.
At the same time for fixed single determinant reference state $|R\rangle $ its
equivalence class  representatives can be uniquely labelled by subsets
$Z\supset  R$, or, in orbital representation, by pairs $(Z_{\alpha},Z_{\beta})
\supset (R_{\alpha},R_{\beta})$ such that $|Z_{\alpha}|+|Z_{\beta}|=p+q$.

For the LiH  ground state ($R_{LiH}=1.595$ \AA ) in the basis of 6 MOs
the HF code is $|HF\rangle =(110000,110000)$ and its equivalence class
for $q=2$ contains exactly ${{12-4}\choose 2}=28$ elements. These elements
$(Z_{\alpha},Z_{\beta})$ are of three different types: type (1,1) of pairs
$|Z_{\alpha}|=|Z_{\beta}|=3$, type (2,0) of pairs $|Z_{\alpha}|=4,|Z_{\beta}|=2$,
and type (0,2) of pairs $|Z_{\alpha}|=2,|Z_{\beta}|=4$. For each representative
of the HF equivalence class we selected the lowest eigenstate of the operator
(48) as a germ generating simple (4,2)-sheaf. Of 28 simple sheaves only 9 proved
to be linearly independent (most probably due to high point symmetry) and the
equivalence class energy  was calculated to give
$E=-7.972047$ a.u. The corresponding wavefunction is not stable and
after its disassembling 35 linearly independent simple sheaves were generated
to give finally stable wavefunction of the CI index 35 with energy
$E=-7.972323$ a.u. Conventional approach leads to CI space of 92 excited
determinants and the  energy value  $E=-7.972323$ a.u. for the lowest CI
vector in this space.

For the water molecule ground state ($R_{OH}=0.957$ \AA ,
$\widehat {HOH}=104.3^{\circ}$)  in the basis of 7 MOs the HF reference code is
$|HF\rangle =(1111100,1111100)$. Its equivalence class contains
${{14-10}\choose 2}=6$ elements of types (1,1),(2,0),and (0,2). The corresponding
codes and energies are:

\begin{tabular}{cccc}
Pair $(Z_{\alpha},Z_{\beta})$ &Energy          &Energy                  & Energy            \\
                              &of HF class     &of HF class before      & of HF class after    \\
                              &representative  &stabilization           & stabilization      \\
                              & (a.u.)         & (a.u.)                 &  (a.u.)            \\
$\left. {\begin{array}{c} 1111110\\ 1111110 \end{array} }\right \}$ & -75.693408   & &\\
$\left. { \begin{array}{c} 1111110\\ 1111101 \end{array} }\right \}$ & -75.687226  & &\\
$\left. {  \begin{array}{c} 1111101\\ 1111110 \end{array} }\right \}$ & -75.687226 & \begin{tabular}[t]{c}\phantom {} \\-75.716895 \end{tabular}&  \begin{tabular}[t]{c}\phantom {} \\-75.728024 \end{tabular} \\
$\left. {  \begin{array}{c} 1111101\\ 1111101 \end{array} }\right \}$ & -75.695390 & &\\
$\left. {  \begin{array}{c} 1111111\\ 1111100 \end{array} }\right \}$ & -75.680388 & &\\
$\left. {  \begin{array}{c} 1111100\\ 1111111 \end{array} }\right \}$ & -75.680388 & &
\end{tabular}

The CI wavefunction involing HF determinant and all single and double excitations
from this determinant as a reference state (on the whole 141 determinants) corresponds
to the energy $E=-75.728063$ a.u. In our calculations we selected the lowest
2-electron wavefunctions for each representative of the HF class by
diagonalizing operator (48): for (1,1) type elements the order of the eigenvalue
problem was  36 (four matrices), for (2,0) and (0,2) type elements we had two
matrices of order 21. All six simple (10,2)-sheaves proved to be linearly
independent and after solving $6\times 6$ CI problem we obtained wavefunction
(sheaf) of the CI index 45. Final stabilization of this wavefunction was
performed in the CI space of dimension 45.

In Tables 1 - 2 integral characteristics of ground state wave functions
of $LiH$ and $H_2O$ molecules obtained by CI method accounting all single
and double excitations from the HF reference state $({\Psi}_{sd})$ and by
FCI method are presented. In these tables term 'actual expansion length'
stands for the numbers of expansion coefficients greater than $10^{-5}$
by absolute value.

The examples discussed are certainly not very impressing and their only
advantage is in their simplicity. But even on these primitive examples
it was possible to
learn what computational difficulties may occur in the course of
our approach implementation to treat larger systems. The first difficulty is
connected with  binary codes (index subsets)  generation. Standard
generation in the lexical order  is known to be slow \cite {Comb}.
Much faster Gray
algorithms \cite {Comb}  should certainly be employed in future implementations.
Technique to determine code number if the binary code is given, and restore code by its
number in the list of codes (without scanning the list) is also required.
The second difficulty is
the selection of linearly independent simple sheaves and their orthogonalization.
The problem here may be connected with large amount of required simple sheaves and
their numerical orthogonalization.  An attempt should be made to develop
some analytic technique for selection of linearly independent simple sheaves.
And at last it is
necessary to have fast routine for the Hamiltonian
matrix elements evaluation in the basis of sheaves obtained by
orthogonalization of simple sheaves.

\newpage

\bigbreak
\centerline{\bf 6. Conclusion}

Our approach to treatment of correlation effects based on new models of the
FCI spaces  opens  new perspective directions  both in mathematical and
computational chemistry. Staying on the zero-level theory (HF equivalence class)
it is  possible to improve the quality of the wavefunctions by increasing
the order of density matrices moving from $(p,2)$-sheaves  to sheaves of
higher order. On the other hand, it is possible to perform calculations in
the Fock space model ${\cal S}_{n,p,q}$ with fixed $q$ and try to reach
the FCI limit either by turning to the 1st, 2nd and higher order theories
or use from the very beginning purely algorithmic approach close in its
idea to the Davidson diagonalization scheme \cite {Davidson}.
There are many things
to be done in the frameworks of our approach both in pure theory and in
practical implementation  of new computational schemes.
Sheaves properties such as their point and spin symmetry, definition of
excited sheaves, etc., are still to be studied. Of special interest are
$(p,1)$-sheaves and their possible connection with the density functional
theory approach. Vector and integral characteristics of
formal many electron wavefunctions obtained in this article are very
intriguing as well.

\bigbreak

\newpage
\centerline{\bf Appendix A. }
\bigbreak

Operator $A^{-1}(n,p,q)$  involved in general expression  (42) for the CI matrix
elements  is defined by its matrix representation
with respect to the basis of $q$-electron determinant generators (18)
and due to the huge dimension of space where it acts it may be very difficult to
perform its direct transformation to some other, probably more convenient, basis.
Therefore, at present stage of theory, all required
matrix elements should finally be expressed via  matrix elements of
Hamiltonian between  determinant $(p,q)$-sheaves. For two determinant sheaves
(45) corresponding to determinant labels $R$ and $R'$ in spin-orbital
representation we have
$$H_{RR'}=
\frac{1}{{{n-p}\choose q}}\sum\limits_{Z\supset R\cup R'}^{(p+q)}
(-1)^{|R\cap {\Delta}_Z|+|R'\cap {\Delta}_Z|}
\langle Z\backslash R|A^{-1}(n,p,q)H_{p\to q}|
Z\backslash R'\rangle\eqno(A.1)$$
In the majority of cases all molecular calculations are performed  in
one-electron bases of molecular orbitals. Turning to MO basis in Eq.(A.1) leads
to more cumbersome expressions but simplifies essentially concrete
calculations.

Using Handy's split representation of spin-orbital index sets \cite {Handy},
it is possible to  replace
$R$ by $(R_{\alpha},R_{\beta})$ and $Z$ by  $(Z_{\alpha},Z_{\beta})$ where
$R_{\sigma}$ and $Z_{\sigma}$ are taken from  the same orbital index set $M$, $|M|=m$
for $\sigma =\alpha ,\beta.$ With fixed projection $M_S$ of the total spin
the numbers $p_{\alpha}$ of ${\alpha}$ and $p_{\beta}$ of ${\beta}$ electrons
are fixed for each basis determinant and the
following simple equalities keep true:
$$|Z_{\alpha}\backslash R_{\alpha}|+|Z_{\beta}\backslash R_{\beta}|=q\eqno(A.2)$$
$$|Z_{\alpha}|+|Z_{\beta}|=p+q\eqno(A.3)$$

Let us start with the analysis of the sign prefactors in Eq.(A.1). To this end
it is convenient to introduce the following notations:
$$K_{\sigma}=R'_{\sigma}\backslash (R_{\sigma}\cap R'_{\sigma})=
\{r_1,r_2,\ldots , r_{k_{\sigma}}\}\eqno(A.4)$$
$$K'_{\sigma}=R_{\sigma}\backslash (R_{\sigma}\cap R'_{\sigma})=
\{s_1,s_2,\ldots ,s_{k_{\sigma}}\}\eqno(A.5)$$
where $|K_{\sigma}|=|K'_{\sigma}|={k_{\sigma}}.$

Technique for handling sign prefactors developed in our previous
article \cite {Panin-1}  makes it easy to get the following  comparison modulo 2:
$$|R\cap {\Delta}_Z|+|R'\cap {\Delta}_Z|\equiv $$
$$\equiv |(R_{\alpha}\Delta R'_{\alpha})\cap {\Delta}_{Z_{\alpha}}|+
|(R_{\beta}\Delta R'_{\beta})\cap {\Delta}_{Z_{\beta}}|+
|(R_{\alpha}\Delta R'_{\alpha}|\times {\rm mod}(|Z_{\beta}|,2)\eqno(A.6)$$

Let us put
$$Z_{\sigma}=R_{\sigma}\cup R'_{\sigma}\cup X_{\sigma}\eqno(A.7)$$
where $ X_{\sigma}$ is a ``free" part of the summation index $Z_{\sigma}$
in Eq.(A.1) and
$$|X_{\alpha}|+|X_{\beta}|=q-(k_{\alpha}+k_{\beta})\eqno(A.8)$$

Now the sign prefactor power can be separated into two parts:
Non-active part that is not involved in summation
$$\varepsilon=\sum\limits_{\sigma}\left [\frac{|R_{\sigma}\Delta R'_{\sigma}|+1}
{2}\right ]+
\sum\limits_{\sigma}|(R_{\sigma}\Delta R'_{\sigma})\cap
{\Delta}_{R_{\sigma}\cap R'_{\sigma}}|\eqno(A.9)$$
and active part
$$a=\sum\limits_{\sigma}|(R_{\sigma}\Delta R'_{\sigma})\cap {\Delta}_{X_{\sigma}}|
+|R_{\alpha}\Delta R'_{\alpha}|\times {\rm mod}(|R_{\beta}\cup R'_{\beta}\cup X_{\beta}|,2)
\eqno(A.10)$$
Brackets in Eq.(A.9) stand for the entier function extracting integral part of its
argument.

The comparision
$$|R_{\sigma}\Delta R'_{\sigma}|=2k_{\sigma}\equiv 0({\rm mod}\quad 2)$$
leads to the following expressions that hardly admit further
simplification
$$\varepsilon=k_{\alpha}+k_{\beta}+
\sum\limits_{\sigma}|(R_{\sigma}\Delta R'_{\sigma})\cap
{\Delta}_{R_{\sigma}\cap R'_{\sigma}}|\eqno(A.11)$$
$$a=\sum\limits_{\sigma}|(R_{\sigma}\Delta R'_{\sigma})\cap {\Delta}_{X_{\sigma}}|
\eqno(A.12)$$

Now it is possible to rewrite  general expression (A.1) for the CI matrix elements
in terms of orbital index sets:
$$H_{(R_{\alpha},R_{\beta}),(R'_{\alpha},R'_{\beta})}=
\frac{(-1)^{\varepsilon}}{{{2m-p}\choose q}}
\sum\limits_{{l_{\alpha},l_{\beta}}\atop {(l_{\alpha}+l_{\beta}=q-(k_{\alpha}+k_{\beta}))}}
\sum\limits_{X_{\alpha}\subset M\backslash (R_{\alpha}\cup R'_{\alpha})}^{(l_{\alpha})}$$
$$\sum\limits_{X_{\beta}\subset M\backslash (R_{\beta}\cup R'_{\beta})}^{(l_{\beta})}
(-1)^a\langle K_{\alpha}\cup X_{\alpha},K_{\beta}\cup X_{\beta}
|A^{-1}(2m,p,q)H_{p\to q}|
K'_{\alpha}\cup X_{\alpha},K'_{\beta}\cup X_{\beta}\rangle\eqno(A.13)$$
Note that for fixed spin projection value both the reduced Hamiltonian
$H_{p\to q}$ and its image $A^{-1}(n,p,q)H_{p\to q}$ are block-diagonal
with exactly $q+1$ blocks.

Eq.(A.13) can be  simplified  for each concrete pair $(k_{\alpha},k_{\beta})$.
In the most important case $q=2$ we have six such pairs:(0,0),(1,0),(0,1),
(1,1),(2,0), and (0,2).

{\bf I.  $(k_{\alpha},k_{\beta})=(0,0)$}

This case corresponds to $R_{\sigma}= R'_{\sigma}$ and
$$H_{(R_{\alpha},R_{\beta}),(R_{\alpha},R_{\beta})}=
\frac{1}{{{2m-p}\choose 2}}
\left [
\sum\limits_{{i<j}\atop {i,j\in M\backslash R_{\alpha}}}
\langle ij,\emptyset |A^{-1}(2m,p,2)H_{p\to 2}| ij,\emptyset\rangle +\right.$$
$$ +
\sum\limits_{{i<j}\atop {i,j\in M\backslash R_{\beta}}}
\langle \emptyset,ij |A^{-1}(2m,p,2)H_{p\to 2}| \emptyset,ij \rangle +$$
$$\left. +
\sum\limits_{i\in M\backslash R_{\alpha}}\sum\limits_{j\in M\backslash R_{\beta}}
\langle i,j |A^{-1}(2m,p,2)H_{p\to 2}| i,j \rangle
\right ]\eqno(A.14)$$

{\bf II.  $(k_{\alpha},k_{\beta})=(1,0)$}

In this case $K_{\alpha}=\{r_1\},K'_{\alpha}=\{s_1\},r_1\ne s_1$,
$\varepsilon=1+|\{r_1,s_1\}\cap {\Delta}_{R_{\alpha}\cap R'_{\alpha}}|$,
$$H_{(R_{\alpha},R_{\beta}),(R'_{\alpha},R_{\beta})}=$$
$$=\frac{(-1)^{\varepsilon}}{{{2m-p}\choose 2}}
\left[
\sum\limits_{i\in M\backslash (R_{\alpha}\cup R'_{\alpha})}(-1)^a
\langle \{r_1,i\},\emptyset |A^{-1}(2m,p,2)H_{p\to 2}| \{s_1,i\},
\emptyset\rangle +\right.$$
$$\left.+\sum\limits_{i\in M\backslash R_{\beta}}
\langle r_1,i |A^{-1}(2m,p,2)H_{p\to 2}| s_1,i\rangle \right], \eqno(A.15)$$
and
$$a=\cases{0,&if $r_1<i$ and $s_1<i$ or $r_1>i$ and $s_1>i$\cr
           1,&if $r_1<i$ and $s_1>i$ or $r_1>i$ and $s_1<i$\cr}\eqno(A.16)$$

{\bf III.  $(k_{\alpha},k_{\beta})=(0,1)$}

In this case $K_{\beta}=\{r_1\},K'_{\beta}=\{s_1\},r_1\ne s_1$,
$\varepsilon=1+|\{r_1,s_1\}\cap {\Delta}_{R_{\beta}\cap R'_{\beta}}|$,
$$H_{(R_{\alpha},R_{\beta}),(R_{\alpha},R'_{\beta})}=$$
$$=\frac{(-1)^{\varepsilon}}{{{2m-p}\choose 2}}
\left[
\sum\limits_{i\in M\backslash (R_{\beta}\cup R'_{\beta})}(-1)^a
\langle \emptyset,\{r_1,i\} |A^{-1}(2m,p,2)H_{p\to 2}| \emptyset,\{s_1,i\}
\rangle +\right.$$
$$\left.+\sum\limits_{i\in M\backslash R_{\alpha}}
\langle i,r_1 |A^{-1}(2m,p,2)H_{p\to 2}| i,s_1\rangle \right], \eqno(A.17)$$
and $a$ is given by Eq.(A.16)

{\bf IV.  $(k_{\alpha},k_{\beta})=(1,1)$}

We have $K_{\alpha}=\{r_1\},K'_{\alpha}=\{s_1\},r_1\ne s_1$,
$K_{\beta}=\{r_2\},K'_{\beta}=\{s_2\},r_2\ne s_2$,
$\varepsilon=|\{r_1,s_1\}\cap {\Delta}_{R_{\alpha}\cap R'_{\alpha}}|
+|\{r_2,s_2\}\cap {\Delta}_{R_{\beta}\cap R'_{\beta}}|$, $a=0$, and
$$H_{(R_{\alpha},R_{\beta}),(R'_{\alpha},R'_{\beta})}=
\frac{(-1)^{\varepsilon}}{{{2m-p}\choose 2}}
\langle r_1,r_2 |A^{-1}(2m,p,2)H_{p\to 2}| s_1,s_2 \rangle\eqno(A.18)$$

{\bf V.  $(k_{\alpha},k_{\beta})=(2,0)$}

In this case $K_{\alpha}=\{r_1,r_2\},K'_{\alpha}=\{s_1,s_2\}$,
$\varepsilon=|\{r_1,r_2,s_1,s_2\}\cap {\Delta}_{R_{\alpha}\cap R'_{\alpha}}|$,
$a=0$, and
$$H_{(R_{\alpha},R_{\beta}),(R'_{\alpha},R_{\beta})}=
\frac{(-1)^{\varepsilon}}{{{2m-p}\choose 2}}
\langle r_1r_2,\emptyset |A^{-1}(2m,p,2)H_{p\to 2}| s_1s_2,\emptyset
\rangle\eqno(A.19)$$

{\bf V.  $(k_{\alpha},k_{\beta})=(0,2)$}

We have $K_{\beta}=\{r_1,r_2\},K'_{\beta}=\{s_1,s_2\}$,
$\varepsilon=|\{r_1,r_2,s_1,s_2\}\cap {\Delta}_{R_{\beta}\cap R'_{\beta}}|$,
$a=0$, and
$$H_{(R_{\alpha},R_{\beta}),(R_{\alpha},R'_{\beta})}=
\frac{(-1)^{\varepsilon}}{{{2m-p}\choose 2}}
\langle \emptyset,r_1r_2 |A^{-1}(2m,p,2)H_{p\to 2}| \emptyset,s_1s_2
\rangle\eqno(A.20)$$

Practicaly all large scale CI calculations are based on the so-called
direct methods \cite {Roos} and use  Davidson-type algorithms for solving partial
eigenvalue problem \cite {Davidson}. In these methods at each Davidson iteration
the result of applying Hamiltonian to the current CI vectors is
calculated and explicit Hamiltonian construction is actually not required.
In the frameworks of our approach it is not difficult to write down general
formula for
Hamiltonian (12) action on some trial $(p,q)$-sheaf $\{{\psi}_Z\}$ (in MSO basis).
We have $$\langle R| H\Psi \rangle =\frac{1}{{{n-p}\choose q}}\sum\limits_{Z\supset R}^{(p+q)}
(-1)^{|R\cap {\Delta}_Z|}\langle Z\backslash R|A^{-1}(n,p,q)H_{p\to q}{\psi}_Z
\eqno(A.21)$$
where $|R\rangle $ is some basis determinant  in  CI
wavefunction expansion and ${\psi }_Z$ are  $q$-germs of $\Psi $ obtained by its
disassembling. This formula, after its rewriting in MO
basis set, can certainly be used in
existing direct CI algorithms but we believe that it is possible to
develop very efficient new methods based on our model of the CI space.
One of such algorithms is briefly outlined in the fourth section.
\newpage

\bigbreak
\bigbreak
\bigbreak

\centerline{\bf Acknowledgment}
\bigbreak
We gratefully acknowledge the Russian Foundation for Basic Research
(Grant 00-03-32943a) and Ministry of Education of RF
(Grant E00-5.0-62) for financial support of the present work. Our
special thanks to Dr. V. F. Bratsev for useful and inspiring discussions.

\newpage
\begin{table}[ht]
\caption{$LiH$ ground state wave functions: Integral characteristics}
\vspace{10mm}
\centering {
\begin{tabular}{ c|c c c c c c }
\hline
  & \multicolumn{3}{ c } {${\Psi}_{sd}$}
         & \multicolumn{3}{ c }  {${\Psi}_{FCI}$} \\
\cline{2-7}
q     & Total    &CI       &Actual   &Total      &CI       &Actual    \\
value & number   &index    &expansion &number     &index   &expansion  \\
      & of germs &         &length    &of germs   &        &length     \\
\hline

 1  &    208     &   35    &25     &   360            &   69  &57     \\
 2  &    482     &   35    &25     &   720            &   69  &52     \\
 3  &    584     &   35    &24     &   752            &   69  &47     \\
 4  &    436     &   35    &10     &   495            &   69  &15     \\
 \hline
 \end{tabular}
 }
\end{table}

\begin{table}[ht]
\caption{$H_2O$ ground state wave functions: Integral characteristics}
\vspace{10mm}
\centering {
\begin{tabular}{ c|c c c c c c }
\hline
  & \multicolumn{3}{ c } {${\Psi}_{sd}$}
         & \multicolumn{3}{ c }  {${\Psi}_{FCI}$} \\
\cline{2-7}
q     & Total    &CI       &Actual   &Total      &CI       &Actual    \\
value & number   &index    &expansion &number     &index   &expansion  \\
      & of germs &         &length    &of germs   &        &length     \\
\hline

 1  &    128     &   49    &32     &   232            &  120  &67   \\
 2  &     69     &   45    &14     &    91            &   61  &21   \\
 3  &     14     &   12    &7      &    14            &   12  &7    \\
 4  &      1     &    1    &1      &     1            &    1  &1    \\
 \hline

 \end{tabular}
 }
 \end{table}
 \end{document}